\documentclass[final,3p,times]{elsarticle}
\usepackage{amssymb}
\usepackage{amsmath}
\usepackage{amsthm}
\usepackage{url}
\usepackage{hyperref}
\usepackage{natbib}
\usepackage{lineno}
\usepackage{deluxetable} 
\usepackage[figuresright]{rotating} 
\usepackage{t1enc}
\usepackage{placeins}

\journal{Journal of Quantitative Spectroscopy and Radiative Transfer}

\begin{document}

\begin{frontmatter}

\title{Spectroscopic study of the light-polluted night sky in Hong Kong}

\author{Chu Wing So} 
\affiliation{organization={Department of Physics, The University of Hong Kong},
            addressline={Pokfulam}, 
            city={Hong Kong},
            country={China}}

\author{Chun Shing Jason Pun} 
\affiliation{organization={Department of Physics, The University of Hong Kong},
            addressline={Pokfulam}, 
            city={Hong Kong},
            country={China}}

\author{Shengjie Liu} 
\affiliation{organization={Department of Physics, The University of Hong Kong},
            addressline={Pokfulam}, 
            city={Hong Kong},
            country={China}}
\affiliation{organization={Spatial Sciences Institute, University of Southern California},
             city={Los Angeles},
             postcode={90089},
             state={CA},
             country={USA}}     

\begin{abstract}
Spectroscopic study of the night sky has been a common way to assess the impacts of artificial light at night at remote astronomical observatories.
However, the spectroscopic properties of the urban night sky remain poorly documented.
We addressed this gap by collecting more than 12,000 zenith sky spectra with compact spectrometers at urban and suburban sites from 2021 to 2023.
Here, by examining the intensity variations of the spectral features that represent characteristic emissions from common artificial light sources, we show that the skyglow is predominantly shaped by artificial emissions, including compact fluorescent lamps and high-pressure sodium lamps.
Contributions from commercially controlled lighting, including those for floodlighting and advertising adopting light-emitting diode and metal halide technologies, were more pronounced in urban areas during the hours leading up to midnight.
We also documented direct evidence of the impact of a neon sign located on top of a commercial tower, illustrating how a single light source can significantly influence the surrounding environment.
Compared with observations made a decade ago at the same location, our findings indicate a growing popularity of light-emitting diode lighting for external use, consistent with the existing literature.
This first comprehensive spectroscopic investigation of light pollution in an urban environment emphasizes the evolving patterns of outdoor lighting and highlights the critical and unique role of spectroscopic measurements.
The results provide essential information for the development of effective strategies and policies to mitigate light pollution in urban areas and at sites of astronomical importance.
\end{abstract}


\end{frontmatter}

\section{Introduction} \label{sec:introduction}
Light pollution -- characterized by excessive and inappropriate use of artificial light at night (ALAN) -- has garnered significant attention due to its detrimental effects on various aspects of life. 
Recent research highlights how light pollution disrupts the natural behaviors and physiological processes of both flora and fauna, leading to negative impacts on ecosystem health~\cite{gaston:2012,holker:2021}. 
Furthermore, it also has significant adverse effects on human well-being and contribute to a variety of health issues~\cite{blume:2019,dbkowska:2023}.
In addition, light pollution robs us of our right to enjoy a natural starry sky and poses a serious threat to professional astronomical observations~\cite{fabio:2016,kyba:2023,falchi:2023,perez:2023}. 

Within the astronomical community, the evaluation of light pollution has generally focused on its photometric effects. This includes assessing how ALAN brightens specific environments, particularly the night skies.
Quantifying the impacts of ALAN has been achieved through ground-based and satellite-based methods. Ground measurements are often conducted using individual devices or networks of photometers, such as the Sky Quality Meter (SQM) and Photometer–Telescope Encoder and Sky Sensor (TESS-W). In contrast, satellite sensors such as the Defense Meteorological Satellite Program’s Operational Linescan System (DMSP-OLS) and the Suomi National Polar-orbiting Partnership satellite's Visible Infrared Imaging Radiometer Suite (VIIRS) provide valuable data from space. For a more in-depth understanding, we recommend consulting recent remote sensing studies on Hong Kong~\cite{liu:2025a,liuhku:2021} and comprehensive reviews available in the literature~\cite{hanela:2018,levin:2020,green:2022,mander:2023,kocifaj:2023d,comino:2023,arroyo:2024}.
Research has consistently provided important information on the geographic and temporal impacts of ALAN on various scales. Based on these findings, the community is dedicated to promoting a sustainable and responsible use of artificial light, with the aim of alleviating the negative effects of light pollution in the long term~\cite{green:2022,morgan-taylor:2023}.

Although photometers and satellite imagery offer convenient and accurate measurements of overall light pollution, they are insufficient to identify the specific sources that contribute to this problem. Photometers are typically panchromatic, responding to the total intensity of incoming light, whereas satellite images are captured within specific spectral bands. This means that measurements from these devices reflect a combined energy output from a variety of outdoor lighting sources alongside natural emissions from atmospheric gases, molecules, and astronomical entities such as the Moon and planets. This amalgamation obscures the individual contributions of different light sources, making it difficult to pinpoint which particular sources are responsible for illuminating the night sky and to understand the mechanisms involved.
As noted in the literature, including studies by~\cite{posch:2018,robles:2021,alejandro:2022,labrousse:2025}, most current photometers and satellite imagery are also insensitive to detecting changes in the emission spectra of ALAN or shifts in the colors of the night sky due to the recent transition to solid-state lighting.

To overcome this limitation, spectroscopy can be utilized. Just as astronomical objects exhibit unique characteristics, each type of ALAN has a distinct spectral signature, referred to as its spectral power distribution (SPD). By analyzing the spectral data of the night sky and decomposing the specific wavelengths emitted by different ALAN sources, we can compare these observations to known SPDs. This comparison enables a more precise identification and quantification of the sources contributing to light pollution.

Traditionally, spectroscopic studies of the night sky have been carried out at professional observatories. Examples in the optical range include the Observatorio del Roque de los Muchachos~\cite{benn:newastro}, Lick Observatory~\cite{osterbrock:1992,slanger:2003}, European Southern Observatory~\cite{hanuschik:2003,cosby:2006}, Paranal Observatory~\cite{patat:2003a,patat:2006,patat:2008}, Xinglong Observatory~\cite{jiang:1999,zhang:2016}, and Calar Alto Observatory~\cite{sanchez:2007}, among others.
While the spectra collected at these locations are primarily dominated by airglow -- characterized by emissions from natural sources such as O$_2$, OI, NO$_2$, and OH -- contaminations from ALAN sources are typically identifiable. These sources include high-pressure sodium, low-pressure sodium, metal halide, and mercury vapor lamps, providing direct evidence of the impact of ALAN on the environment.
Long-term observations, such as those conducted by~\cite{patat:2008} over six years and by~\cite{zhang:2016} over twelve years, have revealed trends regarding the effects of solar activity and ALAN on sky spectra over extended time periods.

In contrast to photometric observations, the spectroscopic study of ALAN and its environmental impacts has been relatively underexplored. This discrepancy is understandable, as spectroscopic observations require more sophisticated and costly equipment, including spectrometers or spectrographs. Additionally, these observations typically require longer integration times and/or larger telescopes to gather data with an acceptable signal-to-noise ratio (S/N). This combination of factors makes spectroscopic studies more difficult to implement compared to their photometric counterparts.

As a practical alternative, researchers have begun measuring the color of the night sky using photometers equipped with filters~\cite{henk:2014,marseille:2021,robles:2021,labrousse:2025}. In addition, calibrated DSLR cameras with fisheye lenses, referred to as Sky Quality Cameras (SQC), have become increasingly popular for assessing light pollution by measuring the correlated color temperature (CCT)~\cite{jechow:2019b,angeloni:2024}. These approaches offer valuable insights while balancing the trade-offs between complexity and cost.
In addition, data obtained from new satellites equipped with multiband sensors, such as SDGSAT-1~\cite{liu:2024b,labrousse:2025} and Jilin-1~\cite{guk:2020}, along with color DSLR images captured by astronauts onboard the International Space Station~\cite{miguel:2019a,miguel:2021a}, have opened up new possibilities for identifying light sources. By decomposing the spectral information into color ratios based on light emissions detected from space, researchers can enhance their understanding of light pollution and its sources.

Facilitated by recent advances in dark-sensitive sensor technology, researchers today are making more effective attempts to observe the sky spectroscopically. The market now offers accurate, portable, and inexpensive spectrometers that have enhanced light sensitivity, making them potentially capable of performing astronomical observations~\cite{kocifaj:2021b}
More sophisticated instruments, such as the SAND
spectrometer~\cite{aube:2016,robles:2021}, meridional
spectrometers~\cite{Taguchi:2002,roldugin:2007}, all-sky
spectrometers~\cite{chernouss:2008,trondsen:2024}, and commerical spectroradiometers~\cite{kollath:2025} have enabled in-depth studies of low-light phenomena. These studies encompass the physics and chemistry of the Earth's upper atmosphere, including investigations into light pollution and auroral behavior. The development of these technologies has significantly broadened the scope of spectroscopic research in the context of environmental impact and atmospheric science.

In this study, we present the first comprehensive spectroscopic analysis of light pollution in the night sky within an urban environment. Using a conventional low-cost spectrometer, initially designed for high-light applications, we conducted measurements at urban locations over multiple years, from 2021 to 2023. 
This effort led to the collection of integrated intensities that reflect characteristic emissions from common ALAN sources, based on more than 12,000 zenith sky spectra, in which about 11,400 spectra were collected at the urban location in 2022-23, while about 600 spectra were collected at a brighter city center in 2021-22. 
Through temporal analysis of these spectra, we were able to identify the ALAN sources contributing to the observed skyglow. 
Additionally, we examined about 500 sky spectra previously obtained a decade ago and from a darker suburban area in 2022.
Our comparative analysis confirmed that the sky was predominantly polluted by emissions from various sources, notably compact fluorescent lamps, high-pressure sodium lamps, and more recently the light-emitting diode. The impact of light-emitting diode and metal halide lighting was particularly pronounced during the early hours of the evening.
We also observed direct evidence of the influence of a neon sign installed atop a commercial tower before midnight. This finding illustrates the dynamic nature of outdoor lighting usage patterns. Our research underscores the importance of spectroscopic measurements in light pollution studies, providing valuable insights to inform mitigation efforts and policies aimed at reducing light pollution. 
The implications of our work extend beyond local observations as many areas worldwide, including astronomical important sites, grapple with the impacts of ALAN.  

The structure of this paper is organized as follows. Section~\ref{sec:methodology} provides an overview of the observational equipment, data quality control measures, statistical analysis of the data, and the analytical methods employed. Section~\ref{sec:result} presents the findings related to line identifications and the temporal variations of ALAN spectral features on different time-scales. Furthermore, this section discusses the impacts of cloud cover and specific light sources on our observations. Lastly, Sections~\ref{sec:discussion} and~\ref{sec:conclusions} summarize our findings and outline potential directions for future research, as well as proposed strategies to mitigate light pollution.

\section{Methods}\label{sec:methodology}

\subsection{Spectrometers \& observing locations}\label{sec:method_spectrometer}

We deployed the \textsc{Ocean Insight} USB2000+ spectrometer (serial no. USB2+U29849) at the observatory located at coordinates $22^\circ ~16'~58.8''$N, $114^\circ ~8'~24.0''$E, within the urban campus of The University of Hong Kong (HKU) in Pokfulam, Hong Kong Island. 
See fig. 1 in~\cite{pun:2014} for the location on the map.
We will refer to this location as Urban 1 from this point onward. 
The spectrometer is equipped with a CCD sensor featuring a 2,048$\times$1 pixel array. Utilizing a 25 $\mu$m slit, it achieves a resolution of 0.34 nm per pixel in the optical range of 340 to 1,028 nm, with a claimed maximum S/N of 250:1.
To ensure durability during long-term deployment, we housed the spectrometer in a metal enclosure fitted with a glass window that caused negligible spectral alteration. The housing provided protection against weather conditions, while the sensor was oriented toward the zenith for optimal sky observations. In our setup, we used the collection lens integrated within the spectrometer to enhance light detection, avoiding the need for fiber optics or additional lenses. The field of view (FoV) of the resulting system is around $30^{\circ}$. The spectrometer was factory calibrated for both wavelength and flux and did not receive direct light from the surrounding environment. We used \textsc{OceanView} software (version 2.0.5), distributed by \textsc{Ocean Insight}, to control the spectrometer during our observations.

We also temporarily installed the spectrometer on the roof of the Hong Kong Space Museum (coordinates 
$22^\circ ~17'~39.0''$N, $114^\circ ~10'~17.2''$E, see fig. 1 in~\cite{pun:2014} for the map) in Tsim Sha Tsui (TST), Kowloon, Hong Kong. TST is located in the heart of the city and serves as a major commercial and tourist hub.
We will refer to this location as Urban 2 from this point onward. 
In addition, we deployed another, more sensitive version of the spectrometer, the \textsc{Ocean Insight} QE Pro (serial no. QEP05349), to the Museum's Sai Kung iObservatory (iObs, coordinates 
$22^\circ ~24'~29.8''$N, $114^\circ ~19'~22.5''$E, see fig. 1 in~\cite{pun:2014} for the map) located at Pak Tam within the Sai Kung West Country Park, New Territories, Hong Kong. 
We will refer to this location as Suburban from this point onward. 
The QE Pro spectrometer features a CCD sensor with a 1,024$\times$1 pixel array. Utilizing a 100 $\mu$m slit, it achieves a resolution of 0.79 nm per pixel across an optical range of 347 to 1,130 nm, with a claimed maximum S/N of 1,000:1. 

From measurements taken with SQM of the \textit{Globe at Night - Sky Brightness Monitoring Network}, the average moonless zenith sky brightnesses were determined to be 16.5, 15.8, and 18.7 arcsec$^{-2}$ at Urban 1, 2, and Suburban, respectively, during the period from 2019 to 2023. These findings provide insight into the conditions of light pollution in these locations, as previously documented in studies by~\cite{pun:2012},~\cite{so:2014}, and~\cite{pun:2014}.

\subsection{Data quality control}\label{sec:method_QC}

We conducted the following data quality control steps on the raw spectroscopic data:

\begin{enumerate}
\item Nonlinearity correction: Non-linearities in the response of the spectrometers were addressed by applying appropriate corrections to the raw spectral data. These corrections were implemented using the factory-provided calibration coefficients for each individual spectrometer. By adjusting the measured intensities based on these coefficients, we ensured that the spectrometers' responses were linearized across the operational range;
\item Sensor saturation removal: Given that data acquisition commenced before sunset and continued past sunrise, some spectra collected during these periods experienced sensor saturation. To ensure the integrity of our data, we discarded any spectrum in which the maximum count approached the saturation level of the spectrometer detector;
\item Twilight removal: We discarded any spectrum that was not acquired during astronomical dark periods. This measure eliminates contamination from twilight effects, which could introduce unwanted variations in the measurements due to residual sunlight;
\item Dark current subtraction: To address the issue of dark current noise inherent in the spectrometers, we regularly acquired master dark frames under conditions similar to those used for our spectral measurements. By removing this unwanted noise component, we improved the signal quality of the spectral data;
\item Transmission checks: We performed regular transmission checks of the glass window using the \textsc{Ocean Insight} HG-2 mercury calibration source to ensure performance. These checks confirmed that the window exhibited a very high transmission efficiency throughout the spectrometer's sensitivity range. Furthermore, we observed that the transmission properties remained stable over time, indicating consistent performance and reliability of the optical system;
\item Sensor calibration: Regular checks of the spectrometers were performed using the HG-2 calibration source to assess and confirm its sensitivity. During these calibration assessments, we found no observable changes in the sensors' sensitivity over time;
\item Spectrometer re-calibration: In August 2023, we returned the USB2000+ spectrometer to the manufacturer for checking and recalibration. Upon evaluation, the manufacturer confirmed that the spectrometer was in good condition, which reassured us of its reliability and performance for ongoing observations; and
\item Moonlight removal: During the analysis of Urban 1 spectra, we implemented a criterion to eliminate any potential interference from moonlight, which could impact our evaluation of ALAN. We discarded any spectra potentially affected by moonlight, specifically those acquired during lunar phases greater than 0.2~\cite{so:2014,pun:2014}.
For Urban 2 observations, previous research by~\cite{so:2014} indicated that moonlight has a negligible impact on the brightness of the sky there. To maximize our dataset and enhance the robustness of our analysis, we opted to include all data collected at Urban 2, regardless of the lunar phase. 
For Suburban, clouds shielded moonlight during the observations. 
\end{enumerate}

\subsection{Sky spectra statistic}\label{sec:method_statistic}

The observations conducted at Urban 1 spanned from 2022 January to 2023 April. Following the data quality control processes outlined earlier, we identified a total of 11,447 spectra that meet the conditions described in Section~\ref{sec:method_QC}. These spectra were collected over 121 distinct evenings, beginning at 20:21 on 2022 January 28 and ending at 05:00 on 2023 March 26 (Hong Kong time, UTC+8, hereafter). 
On average, spectral acquisitions were performed at intervals of 5.4 minutes, with each acquisition comprising an exposure time of $5\times65$ seconds. This method involved averaging five individual spectra per acquisition to enhance signal quality. The nightly sample sizes varied according to seasonal conditions; they ranged from as few as 83 spectra during the shorter night periods in summer to as many as 117 spectra during the extended nights in winter. Additionally, there were instances where hardware or software problems resulted in reduced sample sizes on certain nights. 

The overnight spectroscopic observations conducted at Urban 2 were performed during the evenings of 2021 March 27-28 (with exposure times of $5\times65$ seconds), 2021 July 1-5 (also with exposure times of $5\times65$ seconds), and 2022 March 26-27 (where exposure times ranged from $5\times17$ to $5\times45$ seconds). The primary objective of the March observations was to monitor changes in ALAN emissions in response to Earth Hour light-out events organized by the World Wide Fund for Nature Hong Kong. The findings related to the Earth Hour analysis are detailed in~\cite{so:2025}. A total of 628 spectra that meet the conditions described in Section~\ref{sec:method_QC} were analyzed. Only a small portion (8.8\%) of the spectra was obtained during the one-hour light-out events. Excluding these data from the forthcoming analysis would not yield different results.

At Suburban, data collection took place between between 20:39 and 20:43 on 2022 November 8. For our analysis, we focused on a high-quality spectrum that was captured with an exposure time of $1\times240$ seconds and meets the conditions described in Section~\ref{sec:method_QC}. 

\subsection{Spectral analysis bands}\label{sec:method_band}

\begin{figure}
\centering
\includegraphics[width=\columnwidth]{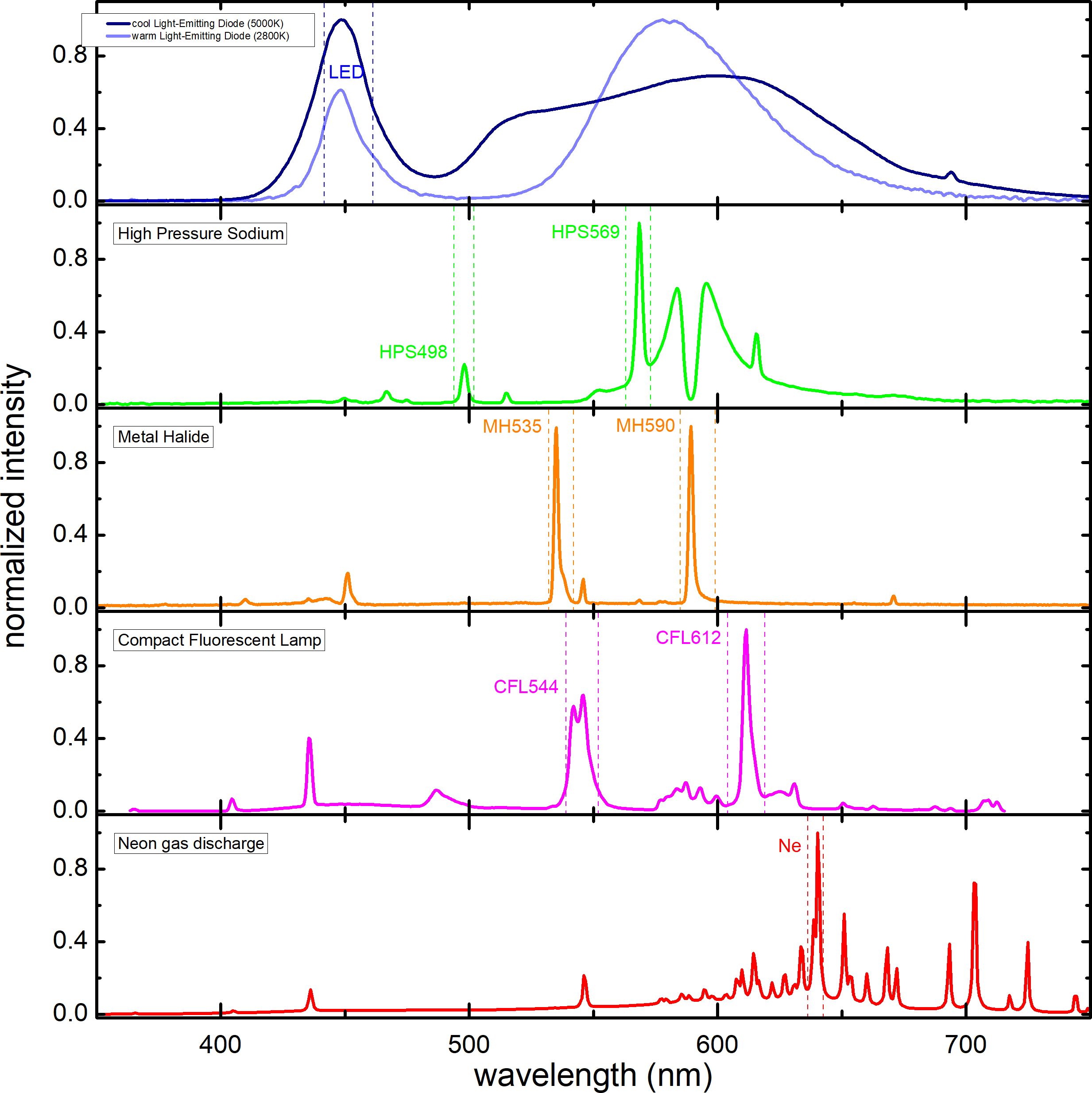}
\caption{Selected SPDs and spectral bands for analysis. The dashed lines indicating the ranges of the chosen bands. The banding is based on the SPDs of common artificial light sources derived from field measurements and online databases. For SPDs displaying multiple representative emissions, such as HPS, MH, and CFL, the central wavelengths are labelled for clarity.\label{fig:spectral_band}}
\end{figure}

To investigate the intensity of ALAN light sources from the sky spectrum, we selected several wavelength regions, referred to as bands. Each band was chosen based on the representative emission characteristics of SPD of common ALAN technologies in Hong Kong (Table~\ref{tab:spectra_banding}): Light-Emitting Diode (LED), High-Pressure Sodium (HPS), and Compact Fluorescent Lamp (CFL).
During our site visits, we observed a transition from traditional metal halide lighting to light-emitting diode floodlights on sports grounds and billboards and a similar decline in the use of gas discharge tubes. To analyse this shift, we also focused on two additional spectral bands: one for traditional Metal Halide (MH) lights and another for Neon gas discharge tubes (Ne). Fig.~\ref{fig:spectral_band} displays the banding overlaid on typical ALAN SPDs.

\begin{table}
\centering
\caption{Selected spectral bands for analysis and their common sources in Hong Kong}
\label{tab:spectra_banding}
\centering
\begin{tabular}{cccc}
\hline 
\textbf{ALAN technology}&\textbf{Band}&\textbf{Wavelength range (nm)}&\textbf{Common sources}\\
\hline 
\textbf{L}ight-\textbf{E}mitting \textbf{D}iode & LED& 442-461& facade lighting, sign boards, TV walls,\\
&&& floodlights for sports grounds and billboards,\\
&&& public area lighting\\
\textbf{H}igh \textbf{P}ressure \textbf{S}odium& HPS& 494-502 \& 563-573& street lighting\\
\textbf{M}etal \textbf{H}alide& MH& 532-542 \& 585-599& floodlights for sports grounds and billboards,\\
&&& public area lighting\\
\textbf{C}ompact \textbf{F}luorescent \textbf{L}amp& CFL& 539-552 \& 604-619& public area lighting\\
\textbf{Ne}on gas discharge& Ne& 636-643& sign boards with gas discharge tubes\\
\hline
\end{tabular}
\end{table}

To avoid confusion, the remainder of this article will use abbreviations (LED, HPS, MH, CFL, and Ne) when referring to the spectral bands or their emissions. The full terms (light-emitting diode, high-pressure sodium, metal halide, compact fluorescent lamp, and neon gas discharge tube) will be used exclusively when discussing specific lighting types.

We evaluated whether variations in the bandwidth centered on the corresponding emission line would influence our results. Our findings demonstrate that the results are not sensitive to minor changes in the bandwidth.

As shown in Fig.~\ref{fig:spectral_band}, the SPD of high-pressure sodium lighting shows a strong, broad, and distinctive emission feature between 573 and 614 nm. This feature overlaps with the pronounced emissions of metal halide lighting at 590 nm (MH590) and compact fluorescent lamps at 612 nm (CFL612). Furthermore, the broad HPS feature is blended with the extensive red component associated with light-emitting diode lighting.
To avoid complications in interpretation due to overlapping signals, we estimate the contribution of high-pressure sodium lamps within two narrow spectral regions characterized by strong peaks at 498 nm (HPS498) and 569 nm (HPS569). This targeted approach enables a more precise assessment compared to the need to rely on broader spectral features.

We determined the wavelength ranges using SPDs from the \textit{Lamp Spectral Power Distribution Database} (LSPDD), which publishes the SPDs of common lighting sources available on the market~\cite{lspdd}. We verified SPDs of light-emitting diodes\footnote{\url{https://lspdd.org/database/2720}, \url{https://lspdd.org/database/2615}}, high pressure sodium\footnote{\url{https://lspdd.org/database/2536}} and compact fluorescent\footnote{\url{https://lspdd.org/database/2510}} lamps against our on-site measurements. However, for metal halide lamps, we found discrepancies and opted to use our own measurements, while we also referenced our data for neon gas discharge lamps since LSPDD does not provide its SPD.

It is important to note that the typical ALAN SPDs selected here represent only a small subset of products and do not cover all ALAN brands, models, and variations. SPDs can vary significantly between different brands and models or due to variations in impurities, even within the same lighting technology. A particularly notable variation is observed with lights of different CCTs, which can exhibit markedly different SPDs~\cite{green:2022,labrousse:2025}.

\subsection{Continuum removal}\label{sec:method_continuum_removal}

Each sky spectrum contains a continuum. To more accurately quantify the intensity of ALAN emissions in Section~\ref{sec:result_R_statistic}, we use the \textsc{Matlab} function \textit{msbackadj}~\cite{matlab_msbackadj} to fit and subtract the signal baseline. This function estimates the continuum with linear interpolation using moving windows of 10-pixel width, spaced at 10-pixel intervals, followed by a smoothing process with a linear fit. 

Reliably determining the continuum spectrum level continues to pose challenges. We also explored non-linear spline interpolation and quadratic smoothing for fitting the continuum, which produced very similar results.

\subsection{Integrating intensity of each band}\label{sec:method_integrating_count}

After subtracting the continuum signal, we calculate the intensity of each band, denoted as \textit{I}, using trapezoidal numerical integration on each continuum-removed spectrum, expressed mathematically as:
\begin{equation}\label{eqt:trapezoidal}
I =\Sigma [y_x + y_{x+1}]\times\Delta x/2
\end{equation}
where $y_x$ and $y_{x+1}$ represent the count values at wavelengths $x$ and $x+1$ respectively, $\Delta x$ is the spectral resolution of the data.

For SPDs featuring multiple narrow emission peaks, such as those from HPS, MH, and CFL, we sum the integrated intensity from each peak associated with the same ALAN species.

\section{Results and Analysis}\label{sec:result}

\subsection{Identification of ALAN spectral features from average spectra}\label{sec:result_typical}

\begin{figure}
\centering
\includegraphics[width=\columnwidth]{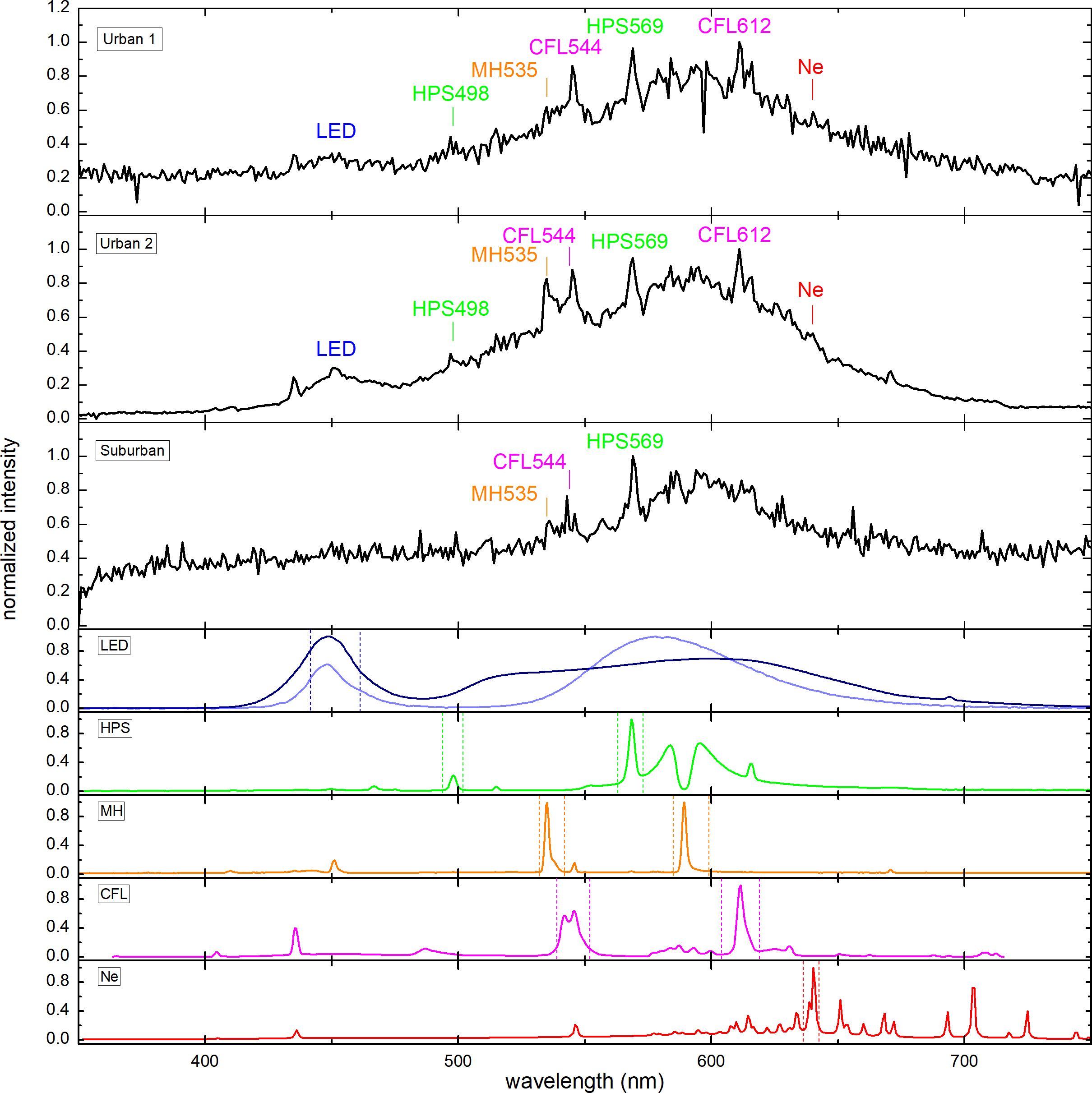}
\caption{
The top three panels display averaged and normalized sunlight-free zenith sky spectra observed at various locations, illustrating the relative contributions of different light usages in comparison to HPS lighting. Major peaks are labelled to align with those identified in Fig.~\ref{fig:spectral_band}. The sky spectra are binned to the nearest integer wavelength.
S/N, as defined in~\cite{stoehr:2008}, are 11.4, 16.9 and 15.8 for Urban 1, Urban 2 and Suburban spectra respectively, for our bands between 440 and 645 nm. 
The bottom five panels, which replicate Fig.~\ref{fig:spectral_band}, present the normalized SPD of common artificial light sources alongside the selected bands. This allows for a comparison between the spectral signatures of the night sky and those emitted by various artificial lighting sources. \label{fig:spectral_typical}}
\end{figure}

Although individual spectra may exhibit noise, line identification becomes feasible when we enhance S/N by averaging all spectra that meet the conditions described in Section~\ref{sec:method_QC}. 
The averaged spectra are shown in Fig.~\ref{fig:spectral_typical}. By matching the emission features of these averaged spectra with those of the reference SPDs within the selected bands indicated at the bottom of the Figure, we can identify the major types of ALAN contributing to the skyglow at each location. Moreover, the averaging process significantly reduces fluctuations caused by varying weather conditions, mainly influenced by cloud cover (see Section~\ref{sec:result_exceptions}).

As seen in the Figure, the Urban 1 and 2 spectra exhibit strong emissions from HPS and CFL, indicating that the observed skyglow is primarily attributed to external lighting utilized for the illumination of public areas and roads. The Urban 2 spectrum, in particular, shows very strong emissions from LEDs and MH, suggesting that floodlights, public area lighting, facade lighting, advertising signboards, TV walls, and similar installations using light-emitting diode and metal halide technologies contribute significantly to brightening the night sky at Urban 2. This observation is not surprising, considering that area serves as a densely populated business and tourist hub.

The analysis of the average urban sky spectra also revealed several weak yet distinguishable emissions. In particular, characteristic Ne emissions were detected in the urban spectra, indicating the presence of neon light sources near the harborfront. A detailed discussion of Ne emissions will be provided in Section~\ref{sec:result_casestudies}.

The Suburban spectrum displays emissions from high-pressure sodium and compact fluorescent lighting, suggesting that the sky brightness at this location is predominantly due to essential lighting sources, such as public area illumination and street lighting. In contrast, the Suburban spectrum does not exhibit strong emissions from light-emitting diode, metal halide, or neon light sources, which are typically employed for non-essential purposes like advertisements and entertainment. This observation leads us to conclude that, in this suburban location, which is situated away from major business districts, the night sky is primarily influenced by essential lighting infrastructure rather than the more widespread use of light-emitting diode, metal halide, and neon lights commonly found in urban areas.

Table~\ref{tab:spectra_typical} provides a summary of the key results of the line identification analysis.

\begin{table}
\centering
\caption{Identification of major ALAN emission features from the average spectra presented in Fig.~\ref{fig:decade} (Urban 1 in 2012) and~\ref{fig:spectral_typical} (all locations in 2021-23). The emissions were categorized based on the quantitative results derived from equation~(\ref{eqt:trapezoidal}), supplemented by visual inspections.} 
\label{tab:spectra_typical}
\centering
\begin{tabular}{ccccc}
\hline
\textbf{Band}&\textbf{Urban 1}      &\textbf{Urban 1}  & \textbf{Urban 2}    &\textbf{Suburban}\\ 
\textbf{Year}& \textbf{2012}        &\textbf{2022-23}  & \textbf{2021-22}    &\textbf{2022}\\ 
\hline 
LED & barely detected      & detected              & clearly detected & barely detected\\ 
HPS & clearly detected     & clearly detected      & clearly detected & clearly detected\\ 
MH  & clearly detected     & detected              & clearly detected & detected \\ 
CFL & clearly detected     & clearly detected      & clearly detected & clearly detected \\ 
Ne  & barely detected      & detected              &  detected        & barely detected\\ 
\hline
\end{tabular}
\end{table}

\subsubsection{Change of spectral composition after midnight}\label{sec:result_typical_earlylate}

\begin{figure}
\centering
\includegraphics[width=\columnwidth]{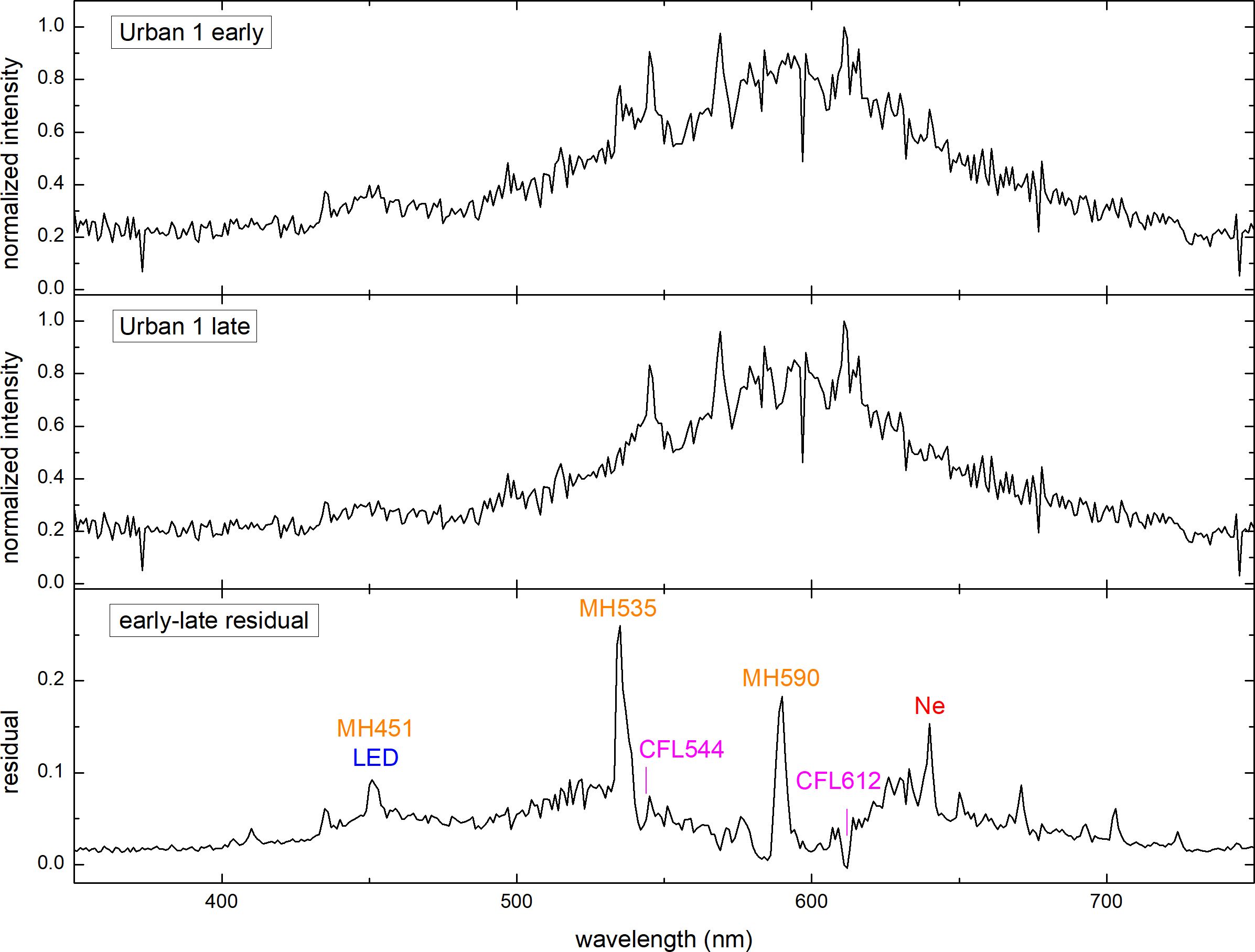}
\caption{Average Urban 1's sky spectra for observations conducted before 23:00 (upper panel, S/N = 11.4) and after 00:00 (middle panel, S/N = 11.5). The bottom panel presents the residual (early-late) in which major peaks are labeled to align with those identified in Fig.~\ref{fig:spectral_band}, signifying changes in the sky's spectral composition after midnight.
Spectra are binned to the nearest integer wavelength. 
\label{fig:spectral_typical_earlylate}}
\end{figure}

In addition to the differences attributed to the variability in light usage between locations, previous photometric studies have indicated that the night sky gradually dims when certain external lighting is turned off as the night progresses~\cite{so:2014,pun:2014}. Therefore, we anticipate that the composition of the sky spectra will also shift as lighting usage evolves throughout the night.
To initiate the temporal study of sky spectra, we divided the average Urban 1 spectrum into two segments: early evening (before 23:00) and late night (after 00:00), considered that the majority of ALAN switched off between 23:00 and 00:00. These segments are presented in Fig.~\ref{fig:spectral_typical_earlylate}. Additionally, we plot the early-late residuals in the same figure to highlight the differences between the two time periods.

The emission peaks for MH and Ne stand out in the residual spectrum, indicating that their contributions to skyglow are significantly stronger before midnight and weaker thereafter. This residual analysis reinforces the conclusion that lighting that utilizes metal halide and neon technologies is a primary contributor to skyglow prior to midnight.
Additionally, a weaker MH emission at 451 nm -- characteristic of SPD -- can be seen in the residual spectrum, although it is not selected as an analysis band (see Fig.~\ref{fig:spectral_band}). This observation further supports our findings. Further discussions on the fading of the Ne band will be provided in Section~\ref{sec:result_casestudies}.

The residual spectrum does not exhibit any emissions from the HPS band, indicating that both the intensity of high-pressure sodium lighting and its usage remain relatively constant throughout the night. Additionally, the residual spectrum does not clearly reveal the broad emission of the LED in the range of 442-461 nm. This lack of distinctiveness is probably due to the overlapping MH emission at 451 nm, which appears in the same spectral range within the residual spectrum.

Finally, with respect to the CFL band, which features two emission peaks at 544 nm and 612 nm, the residual spectrum shows only a weak depression in the latter range (612 nm) and not in the former (544 nm). This observation warrants further investigation to understand the underlying reasons for the differing behavior of the two emission peaks.


\subsection{Quantifying nightly variations of ALAN spectral features}\label{sec:result_R_statistic}

To quantify the spectral dynamics discussed in the previous section, we calculated the intensity $I$ for each band using equation~(\ref{eqt:trapezoidal}), separating the data into two distinct time periods. We define the darkening ratio $R$ as follows:
\begin{equation}\label{eqt:ratio}
R = (I_{\text{late}} - I_{\text{early}})/ I_{\text{early}}\times 100\%
\end{equation}
where $I_{\text{late}}$ represents the averaged intensity for the spectra observed after 00:00 (late night), while $I_{\text{early}}$ refers to the same for the spectra observed before 23:00 (early evening). According to this definition, a positive $R$ indicates that the sky intensity for a particular band was brighter during late night compared to early evening. In contrast, a negative $R$ means that the sky intensity of that band was darker late night than early evening.
Equation~(\ref{eqt:ratio}) is not symmetric, with values ranging from -100 to greater than 100.

Analysis of the darkening ratio $R$ uncovers important general trends, providing valuable information on the patterns of external lighting use and the evolution of lighting infrastructure throughout the night. 

\begin{figure}
\centering
\includegraphics[width=\columnwidth]{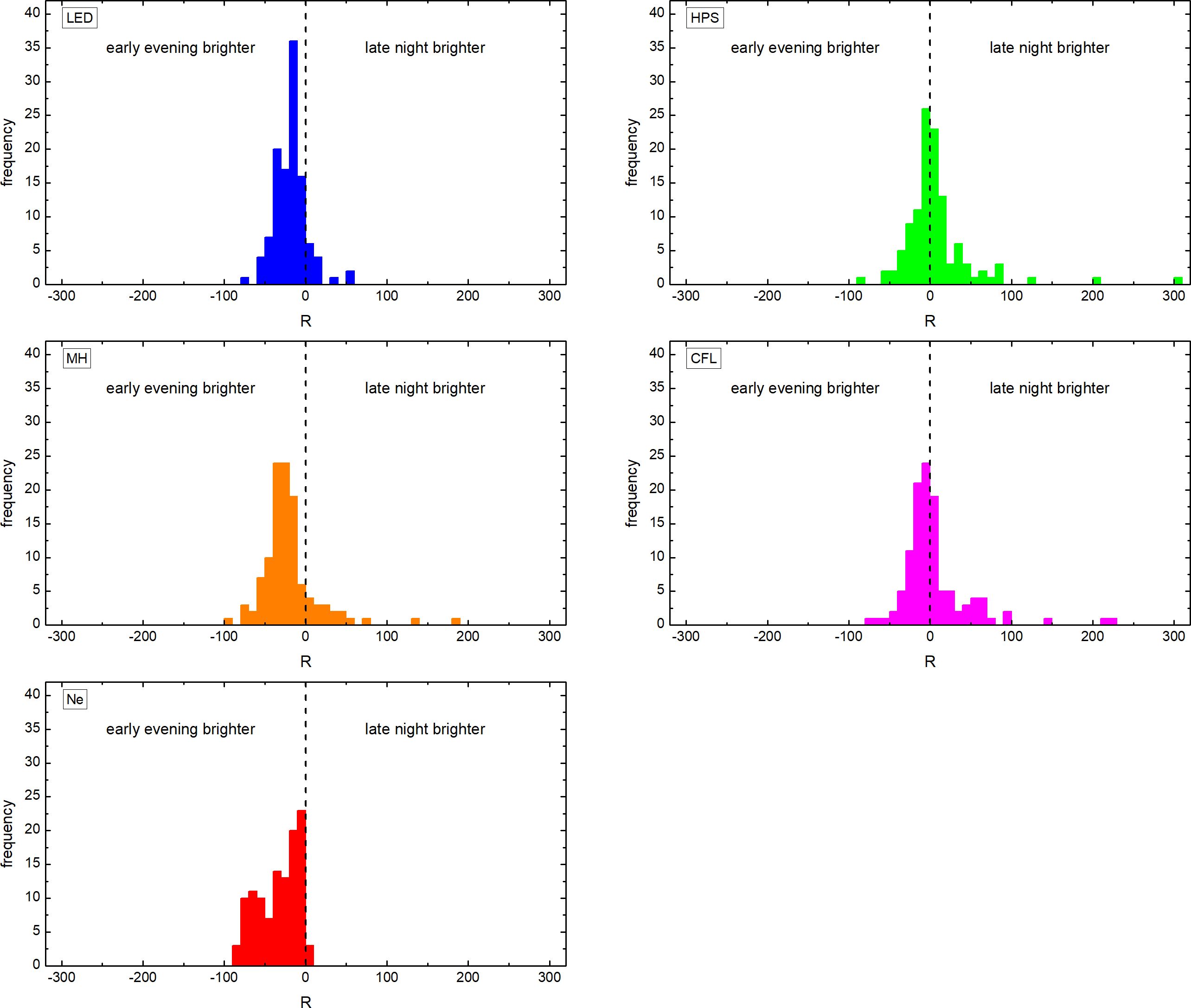}
\caption{Histograms of $R$ in different spectral bands as observed at Urban 1 during moonlight-free period, quantifying nightly variations of ALAN spectral features. The vertical dashed lines represent $R=0$, i.e., same early evening and late night intensities.\label{fig:spectral_ratio_histogram}}
\end{figure}

At Urban 1, we compiled a total of 114 individual data pairs of early and late night measurements for analysis, noting that some nights lacked data for either early evening or late night. Fig.~\ref{fig:spectral_ratio_histogram} illustrates the histograms of the darkening ratio $R$ for each spectral band, providing a visual representation of the distribution and variability of darkening ratios across different types of lighting.
To highlight the central tendency of these distributions, we calculated the median values of $R$ for various lighting types as follows: LED (-16.7), HPS (0.2), MH (-25.9), CFL (-5.3), and Ne (-29.2). These median values are also detailed in Table~\ref{tab:darkening_ratio}.

The data distributions for LED, MH, and Ne exhibit a negative skew, with their median $R$s are smaller, in which $R$s of MH and Ne are particularly smaller ($<$-25) than others.
These observations indicate that the intensities of LED, MH and Ne were significantly dimmer during the late night hours after midnight, and the darkening was more pronounced for MH and Ne.
In contrast, the HPS and CFL data distributions appear to be more symmetric at $R=0$, as indicated by their near-zero values $R$.

We conducted Kolmogorov-Smirnov (K-S) tests to determine if two datasets were drawn from the same distribution~\cite{william:2007}, focusing on the darkening ratio $R$ in relation to HPS. 
The tests showed a P-value of 0 for LED, MH and Ne, indicating significant differences from HPS, while CFL had a P-value of 0.05, suggesting similar distributions with HPS. 
Further KS tests related to MH revealed P-values of 0 for both HPS and CFL, indicating they do not share distributions with MH. In contrast, LED and Ne showed non-zero P-values, suggesting similarities among MH, LED and Ne. Overall, the findings align with the histogram shapes in Fig.~\ref{fig:spectral_ratio_histogram}.

\begin{table}
\centering
\caption{Values of darkening ratios $R$s from different datasets of Urban 1 observations (Section~\ref{sec:result_R_statistic}). The last two columns tabulate two notable exceptions to the general trends, attributed to the significant changes of cloud conditions near midnight coincidentally (Section~\ref{sec:result_exceptions}).
\label{tab:darkening_ratio}}
\centering
\begin{tabular}{cccc}
\hline   
\textbf{Band}&\textbf{2022-23 median}&\textbf{Clear\textrightarrow Overcast Night}&\textbf{Overcast\textrightarrow Cloudy Night}\\
\hline 
LED & -16.7& 55.3& -72.0\\
HPS & 0.2& 309.0& -82.1\\
MH  & -25.9& 186.0& -91.6\\
CFL & -5.3& 216.5& -77.1\\
Ne  & -29.2& -54.8& -69.0\\
\hline
\end{tabular}
\end{table}

These observations further reinforce our interpretations on the use of ALAN derived from the residual analysis discussed in Section~\ref{sec:result_typical_earlylate}. Specifically, the findings indicate that ALAN illumination from light-emitting diode, metal halide, and neon discharge technologies tends to dim after midnight, with the most pronounced reduction noted in neon lighting.
In contrast, high-pressure sodium and compact fluorescent lamps largely maintain consistent illumination levels throughout the night, showing less variability in intensity.

We explored alternative temporal segmentations and found that shifting the cut-off time between 23:00 and 00:00 does not affect the key findings presented above.

\subsection{Effects of clouds}\label{sec:result_exceptions} 

Many studies have highlighted the impact of cloud coverage on observed sky brightness. In particular, it has been universally shown that at light-polluted locations, cloudy skies tend to appear brighter than clear skies due to backscattered ALAN from the ground to the detector~\cite[for
examples]{christopher:2011,kyba:2012,johannes:2014a,johannes:2014b,aube:2016,andreas:2017,bara:2019b,posch:2018,sciezor:2019,sciezor:2020b,karpinska:2023}.
Theoretical radiative transfer models for ALAN propagation in cloudy atmospheres have also reached a mature level of development~\cite{miroslav:2007,kocifaj:2010:lighting,kocifaj:2014}. 

\begin{figure}
\centering
\includegraphics[width=\columnwidth]{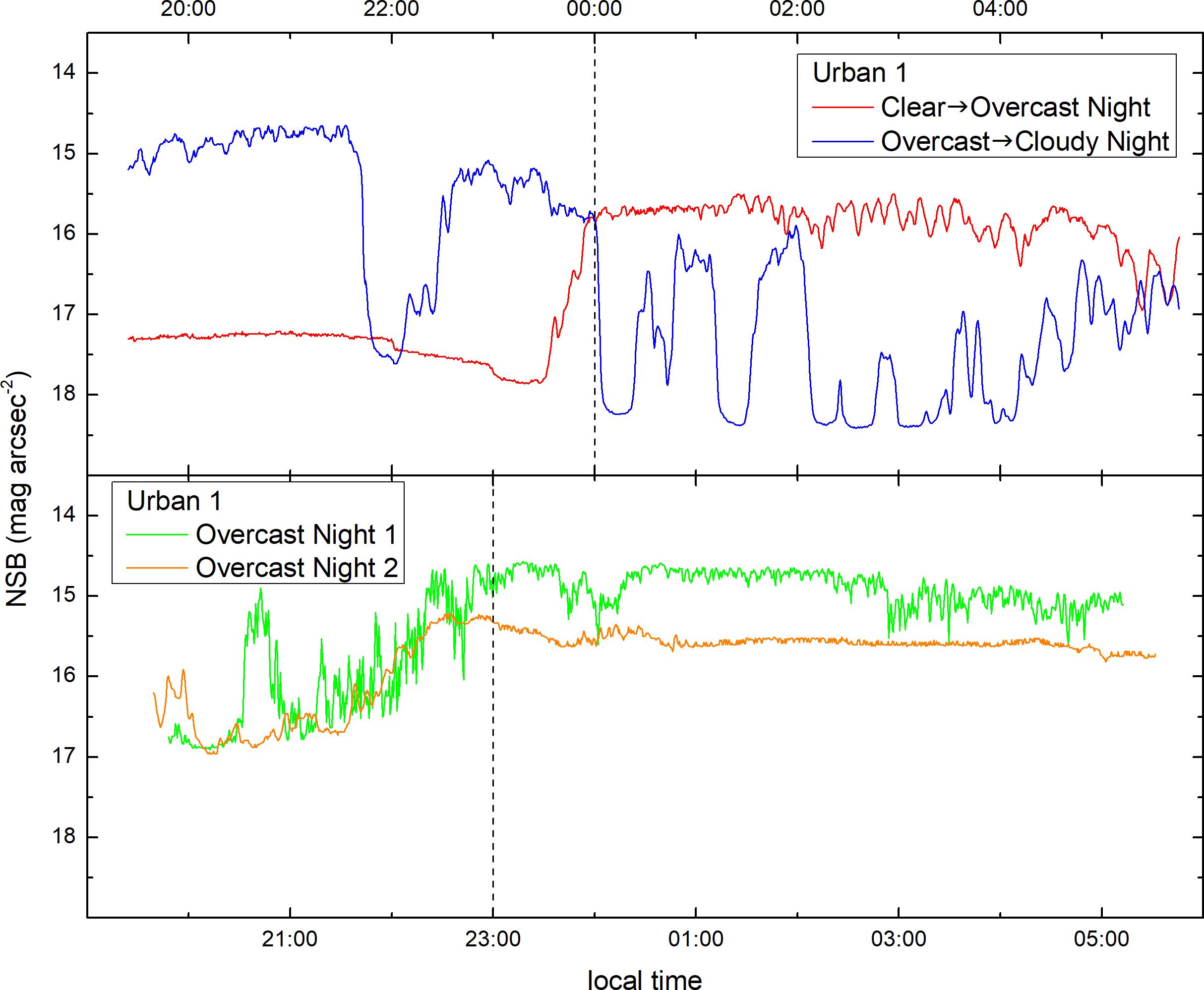}
\caption{Urban 1's NSB light curves in the upper panel illustrate the sharp transitions in cloud coverage at midnight (indicated by the vertical dashed line) for the Clear\textrightarrow Overcast and Overcast\textrightarrow Cloudy Nights. These transitions highlight the significant changes in sky conditions and their subsequent effects on the darkening ratios $R$ (see Section~\ref{sec:result_exceptions}). 
In contrast, the light curves for two Overcast Nights in the bottom panel show relatively little variation in cloud amounts after 23:00 (indicated by another vertical dashed line), enabling us to isolate the impact of artificial lighting from fluctuations in cloud conditions
(see Section~\ref{sec:result_casestudies}).
The NSB measurements were conducted using a SQM-LE at HKU (Urban 1), as part of the \textit{Globe at Night - Sky Brightness Monitoring Network}. These measurements are plotted on the same scale, allowing for easy comparison of relative changes in sky brightness across the different nights. 
\label{fig:nsb_lightcurves}}
\end{figure}

The enhanced backscattering of artificial light from the clouds explains some exceptions to the general trends observed in the darkening ratio $R$ presented in the previous section. Specifically, on the night of 2023 January 24-25, we recorded unusually large $R$ values for all bands except for Ne. A review of the all-sky images and the night sky brightness (NSB) curve for that night, shown in Fig.~\ref{fig:nsb_lightcurves}, revealed that overcast conditions developed after midnight. This shift led to a substantial increase in sky brightness -- approximately 10$\times$. To indicate cloud variation, we refer to this night as Clear\textrightarrow Overcast Night from this point onward.
This dramatic change in atmospheric conditions contributed to the significant spikes in the $R$ values. However, it is important to note that the $R$ value for Ne remained negative, implying that the neon light sources may have already been turned off before midnight, before the cloud cover affected the brightness. Further discussion of this anomaly and its implications will be explored in Section~\ref{sec:result_casestudies}.

On the night of 2023 January 25-26, we observed a reversal in the patterns of the darkening ratio $R$ compared to the previous night. This night exhibited very low $R$ values in all spectral bands, indicating a significant darkening of the sky attributed to cloud thinning occurring after midnight. 
We will refer to this night as Overcast\textrightarrow Cloudy Night from this point onward.
As illustrated in Fig.~\ref{fig:nsb_lightcurves}, the sky appeared darker after midnight, as if all lighting types were effectively inactive during the late night hours.

We have also compiled the extreme values discussed above in Table~\ref{tab:darkening_ratio} for comparison. 

\subsection{In-depth light curve examination}\label{sec:result_casestudies}

\begin{figure}
\centering
\includegraphics[width=\columnwidth]{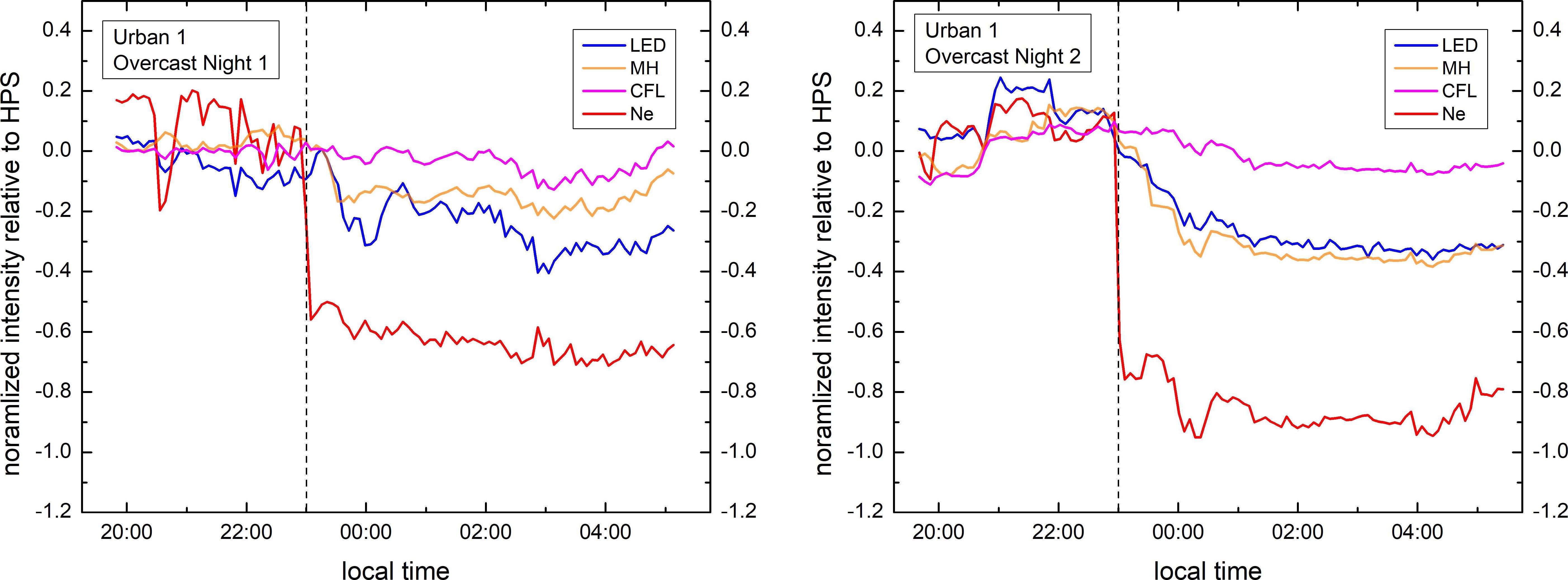}
\caption{Intensities of different bands of the sky spectrum taken during the Overcast Nights 1 and 2 at Urban 1. 
Each light curve presented illustrates the relative change in integrated intensity for each type of ALAN, with HPS serving as the reference point, as it maintained a relatively constant intensity throughout the night (with a
$R$ value of 0.2, as discussed in Section~\ref{sec:result_R_statistic}). The vertical dashed line marks the 23:00 time point, during which a significant drop in Ne intensity is observed, accompanied by a gradual darkening of LED and MH intensities. In contrast, CFL intensity maintains roughly constant brightness throughout the night.
\label{fig:lightcurve_overcast_nights}}
\end{figure}

As illustrated in the previous section, natural factors such as variations in cloud cover can significantly influence the interpretation of darkening ratios. 
To better isolate the impact of artificial lighting from fluctuations in cloud conditions, we concentrated on specific nights that were predominantly overcast, exhibiting relatively consistent cloud amounts, judging from the NSB light curves shown in Fig.~\ref{fig:nsb_lightcurves}. The consistent overcast conditions not only minimize the variability in cloud cover, but also enhance S/N as a result of the increased backscattering of ALAN attributed to the clouds.
Several nights meet the criteria. We present the nights of 2022 March 19-20 (Overcast Night 1) and 2023 February 25-26 (Overcast Night 2) in Fig.~\ref{fig:lightcurve_overcast_nights}, to illustrate this phenomenon.

To account for the wide range of relative intensities across each type of ALAN, we present the normalized intensity in the figures. This normalization allows for clearer comparisons among the different lighting types by adjusting their intensities to a common scale. Furthermore, to further mitigate the influence of cloud variation on our analysis, we depict the relative change in band intensity relative to that of the HPS band, which was observed to maintain stable intensity throughout the night (with a
$R$ value of 0.2, as discussed in Section~\ref{sec:result_R_statistic}). This process involves subtracting the normalized light curve of the HPS band from that of each other band.

During Overcast Night 1, the light curves for the different ALAN bands exhibit fluctuations within a narrow range of approximately $\pm0.2$ before 23:00, until an abrupt drop in Ne intensity occurs at that time. Between 23:00 and 23:30, both the LED and MH bands gradually dim at a similar rate. The LED band continues to exhibit further dimming until midnight, while the Ne band remains at a relatively low level for the remainder of the night.
In contrast, CFL band maintains a relatively bright level throughout the night, highlighting its different operational characteristics compared to the other light types.

Similar trends are evident in the light curves throughout the study period, particularly the observed declines in the intensities of LED, MH, and Ne. These trends are also observable in Overcast Night 2. These two selected nights cover nearly the entire dataset period from 2022 January to 2023 April, suggesting that the observed patterns of intensity fluctuations are consistent over this extended time frame. 

If we assume that the cloud backscattering of ALAN is wavelength independent -- an assumption supported by the Mie scattering properties of cloud water droplets, as noted by~\cite{serrano:2015} -- we can correlate the stability of CFL band and the temporal variations observed in LED, MH, and Ne bands with nightly ALAN usage patterns.
Specifically, while the overall usage of compact fluorescent lamps may have varied, their relative usage compared to high-pressure sodium lighting might have remained stable throughout the night. In contrast, the usage patterns of commercially controlled lighting technologies, such as light-emitting diode, metal halide, and neon discharge, were significantly more pronounced in the hours leading up to midnight, reflecting a peak in artificial lighting demand during that time.
These findings offer insights into the early-late residual spectrum described in Section~\ref{sec:result_typical} and help explain the negatively skewed data distribution of the darkening ratio discussed in Section~\ref{sec:result_R_statistic}. 

\subsubsection{Pinpointing {Ne} light source}\label{sec:ne_source}

Examination of the light curves indicates that significant changes in relative intensity frequently occur at the start of the hour or half-hour marks, particularly at 23:00, 23:30, and 00:00. This pattern suggests a potential relationship between these dimming phenomena and the operational schedules of various types of ALAN. Notably, the abrupt drop in Ne intensity observed at 23:00 prompts us to associate this occurrence with the operational characteristics of neon signs, especially the red variants, which are known to dim regularly and have observable effects on the sky above our observation site. 

Our investigation commenced with the confirmation that the observed red light did not originate from the university campus or the nearby residential neighborhood. We proceeded to analyse the all-sky images captured at Urban 1, which revealed a noticeable dimming of the northeast skyglow around midnight. This direction encompasses business districts such as Sheung Wan.
Upon reviewing maps and conducting site visits in the area, we identified a prominent group of bright neon signs located on top of the Shun Tak Centre West Tower, approximately 1.3 km from our observation site. The billboard, henceforth referred to as STB (Shun Tak Billboard), features a design with four sides, each displaying the tower's name in both English and Chinese. The text is presented in white against a vivid red background, which contributes significantly to the observed skyglow.

\begin{figure}
\centering
\includegraphics[width=\columnwidth]{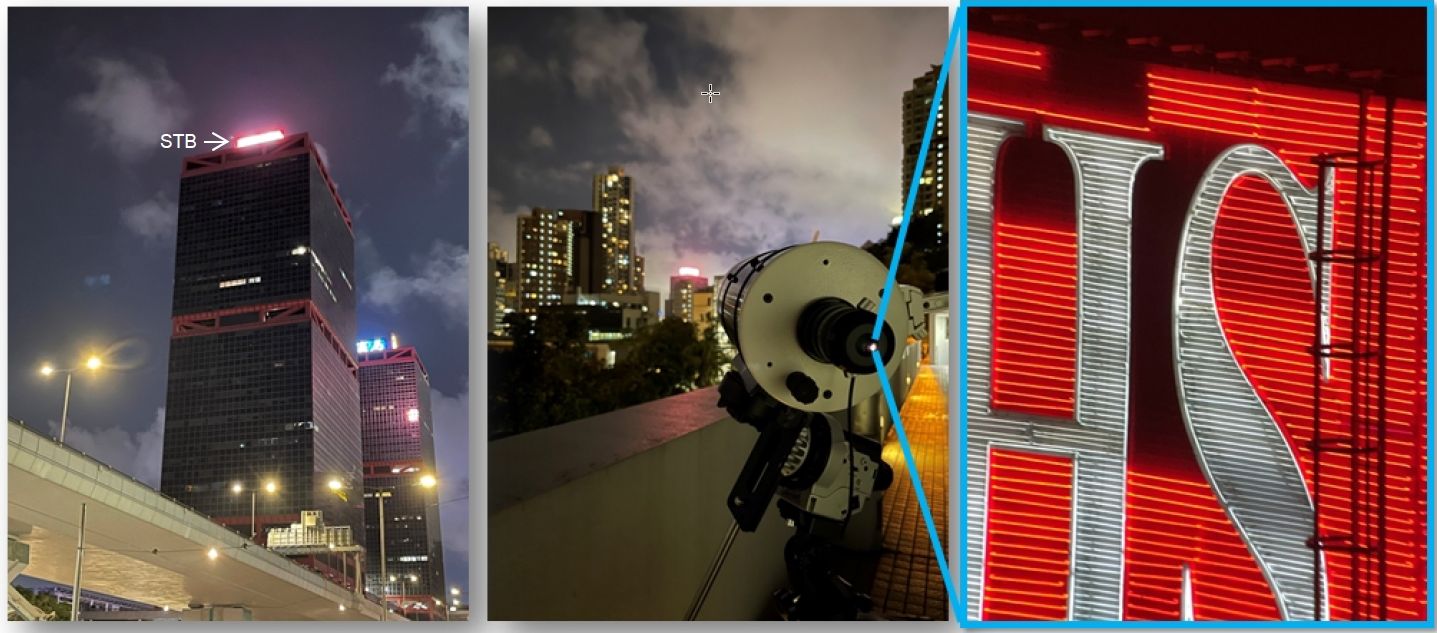}
\caption{STB as seen from street (left) and telescope (right, mirrored image). The close-up image reveals individual neon discharge tubes clearly.
\label{fig:STB_telescope}}
\end{figure}

\begin{figure}
\centering
\includegraphics[width=\columnwidth]{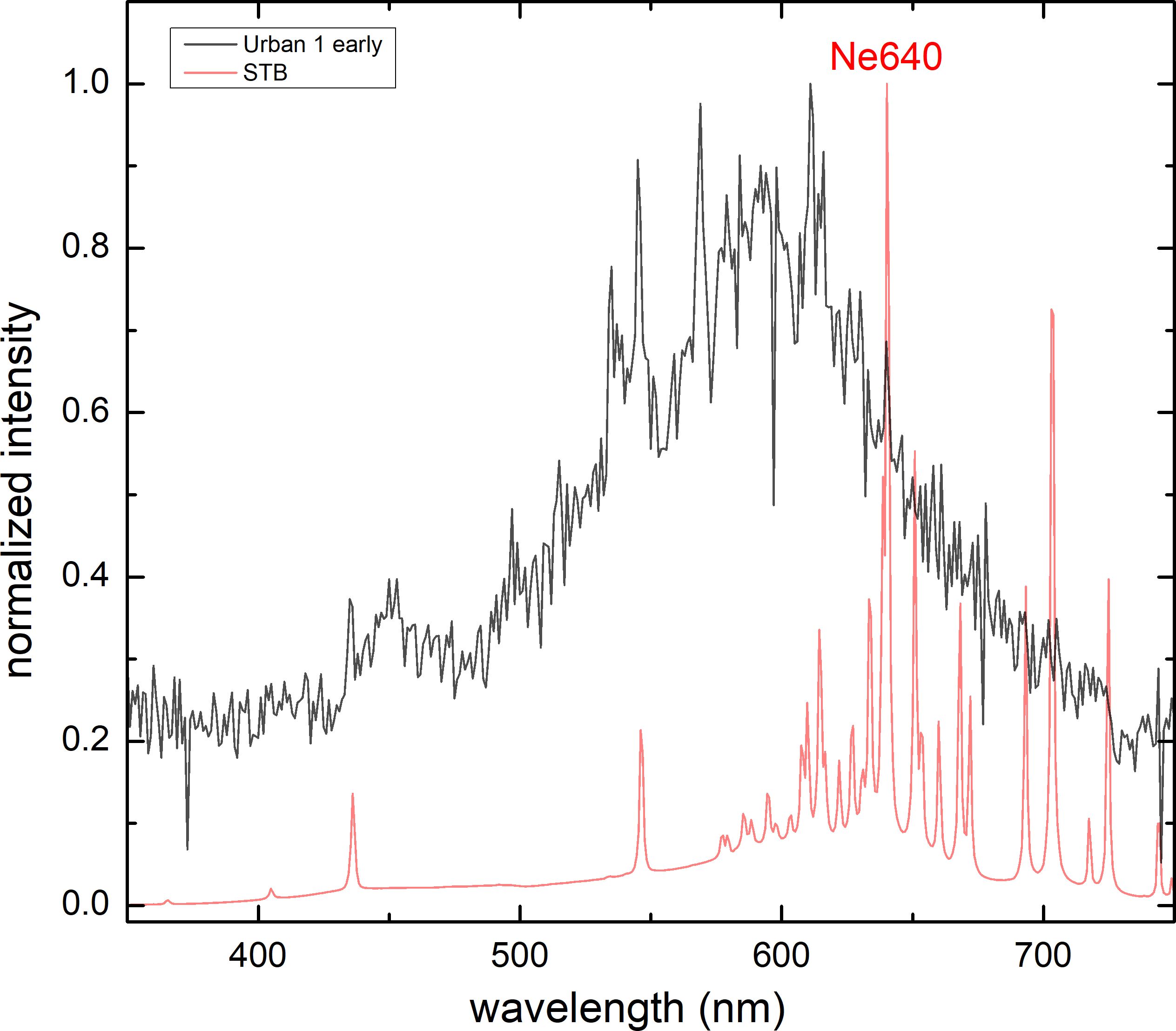}
\caption{The spectrum from STB is presented as a red curve, overlaid on top of the average Urban 1's early evening sky spectrum, which is shown as a black curve and same as that in Fig.~\ref{fig:spectral_typical_earlylate}. Several neon emission lines can be observed within the sky spectrum, including the strongest emission line at 640 nm (labelled as Ne640), corresponding to the prominent emission from the STB. The sky spectrum is binned to the nearest integer wavelength.\label{fig:STB_spectrum}}
\end{figure}

In May 2024, we measured the spectrum from the side of STB that faces Urban 1 using a telescope, as shown in Fig.~\ref{fig:STB_telescope}, in conjunction with a QE Pro spectrometer. The results of this spectral analysis are presented in Fig.~\ref{fig:STB_spectrum} and reveal a strong and characteristic emission line at 640.3 nm (labelled as Ne640), which is associated with the neon gas discharge tubes used in the billboard lighting.
This prominent emission line corresponds to the Ne band and aligns with an emission observed in the sky spectrum captured at Urban 1. Additionally, the spectral analysis identified several weaker emissions from the neon tubes within the overall sky spectrum, further reinforcing the connection between the STB and the observed skyglow.

After establishing the Ne source from STB, we investigated its operational schedule using weather photos captured by the Hong Kong Observatory. These photos, taken at five-minute intervals, revealed that the STB was activated daily between 18:05 and 18:10, during twilight, so sky spectra from this period were excluded from our analysis. 
The billboard was switched off between 23:05 and 23:10 each night, during which a significant drop in Ne intensity is observed.

The combination of spectral evidence and operational timing presents a strong case for the causal link between the operation of STB and the observed presence and darkening trends of Ne emissions in the night sky. In other words, the established usage patterns of the STB throughout the night directly contribute to the changes observed in the sky spectra. This finding emphasizes that even a single bright artificial light source can significantly degrade the surrounding environment, affecting sky brightness from a considerable distance. In addition, the detection of Ne emissions within the urban spectra suggests that some neon-based lighting systems continue to operate within the study area. This persistence occurs despite the general trend towards the adoption of more efficient lighting technologies in the city, as highlighted in~\cite{neon_hktb}. 

Unlike neon-based lighting, light-emitting diode and metal-halide lighting are prevalent in applications such as floodlighting for sports grounds, outdoor billboards, and public lighting. This widespread deployment complicates efforts to identify specific lighting fixtures that are responsible for the observed gradual darkening trends in the night sky.
The gradual nature of these trends indicates that the emissions from light-emitting diode and metal halide lighting in the sky spectra likely arise from multiple independent groups of fixtures. These fixtures differ in characteristics and operate on various lighting schedules, either dimming down or turning off at some times or remaining on throughout the night.
Each group may contribute to the overall light pollution and observed changes in the sky spectral compositions, making it challenging to isolate the influence of individual fixtures or lighting types.

\subsection{Sky spectra over a decade}\label{sec:result_decade}

\begin{figure}
\centering
\includegraphics[width=\columnwidth]{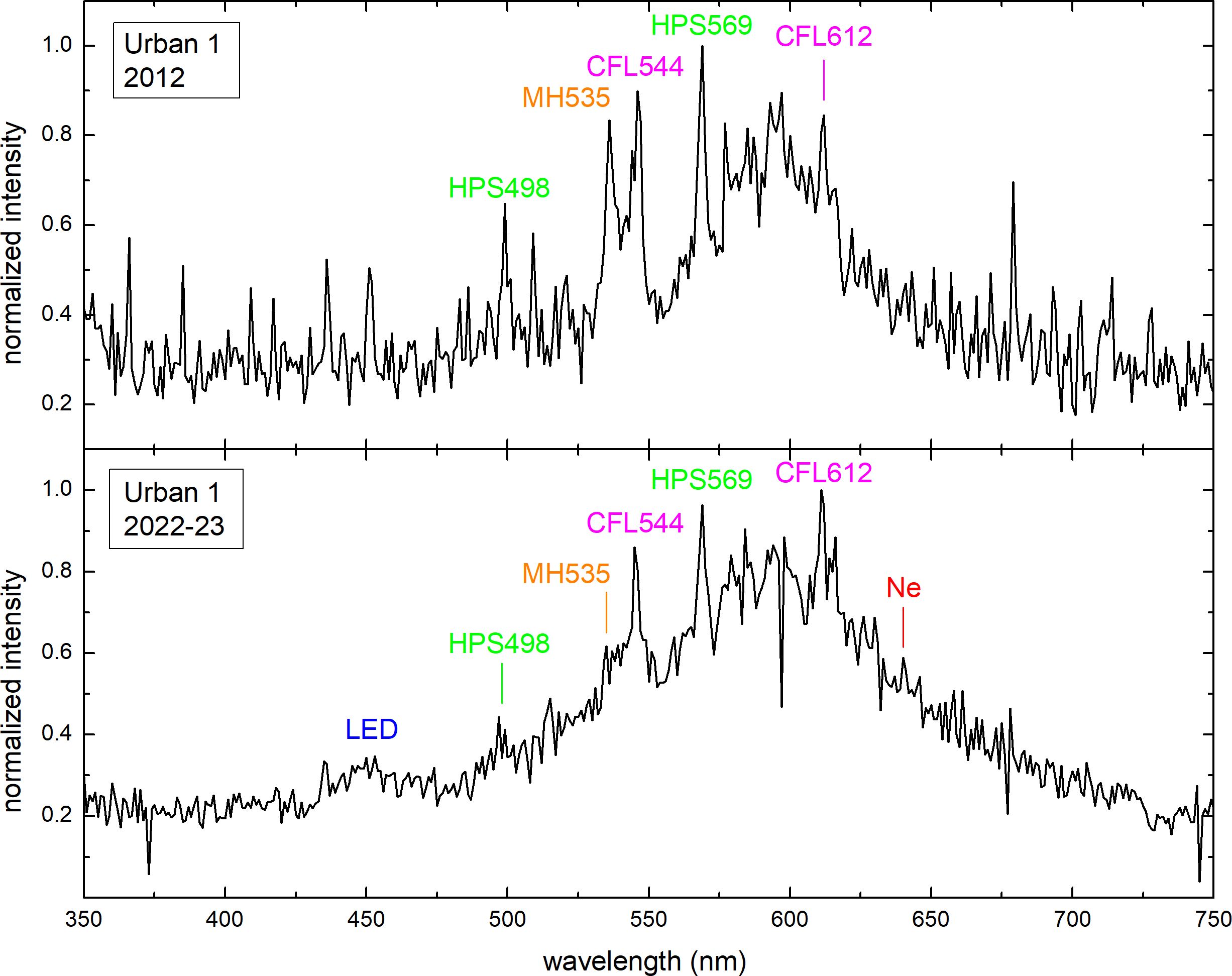}
\caption{Urban 1's 2012 (upper panel, S/N = 3.3) and 2022-23 (bottom panel, same as that in Fig.~\ref{fig:spectral_typical}) average sky spectra, binned to the nearest integer wavelength. 
Major peaks are labelled to align with those identified in Fig.~\ref{fig:spectral_band}, signifying changes in the sky's spectral composition over the past decade.
\label{fig:decade}}
\end{figure}

In 2012, we conducted observations of the night sky at Urban 1 using the same model spectrometer employed in our current investigations. The methodology for the 2012 observations was basically the same as that in 2022-23, except that the housing was not adopted (Section~\ref{sec:method_spectrometer}). During this period, multiple observations were made between late March and early April, covering five individual nights, resulting in a total of 501 sunlight-free spectra in which each acquisition comprising an exposure time of $5\times65$ seconds. 
Following the data reduction processes described in Section~\ref{sec:methodology}, we generated an average spectrum for 2012, which we now present alongside the average Urban 1 spectrum obtained during 2022-23 for comparison in Fig.~\ref{fig:decade}. 
The 2012 spectrum is more noisy primarily due to a much smaller averaging sample size (501 in 2012 v.s. 11,447 in 2022-23). 
To enhance the detection of spectral features, we decided to include the 2012 spectra captured even on moonlit nights. This period included the transition from the first quarter Moon to the full Moon; however, cloud cover obstructed lunar visibility, as indicated by the observed NSB light curves.

The decade-long comparison illustrated in Fig.~\ref{fig:decade} demonstrates that emissions associated with high-pressure sodium and compact fluorescent lamp lighting have been present since at least 2012 and continue to be observed in the 2022-23 spectra. This stability indicates the sustained use of these lighting technologies in the urban landscape over the past decade.
In contrast, the detection of LED emissions in the range of 442-461 nm appears to be a new addition to the spectral profile observed in 2022-23. This suggests that the transition towards light-emitting diode technology is actively taking place, reflecting a shift in lighting practices and technology in the urban environment.
The presence of Ne emission in the noisy spectra from 2012 observations is considered doubtful.

Another significant observation from our decade-long spectral analysis is the decline in emissions near MH band, which were notably strong in the 2012 dataset but have diminished by 2022-23. As illustrated in Fig.~\ref{fig:spectral_typical_earlylate}, the MH emissions were markedly stronger during the early evening hours compared to late night in more recent observations.
Combining these observations, we conclude that the use of metal halide for external floodlighting was more prevalent in 2012, likely being switched on throughout the night. Although some metal halide lights were still in use after a decade, their use was primarily limited to hours before midnight.


\section{Discussion}\label{sec:discussion}

Aside from site surveys conducted at professional observatories, spectroscopic studies of the night sky are notably scarce. With only a few significant exceptions, the spectral composition of light-polluted skyglow has been subject to limited investigation, probably due to historical instrumental constraints. However, given the growing environmental and biological concerns surrounding ALAN and its impacts as communities transition to new lighting technologies, there is an urgent need to explore the spectral components of skyglow.

In this study, we addressed the critical gap in our understanding of urban light pollution by employing conventional, low-cost spectrometers. Our findings demonstrate that these instruments are capable of effectively recording light-polluted sky spectra with reasonable S/N, even without the use of telescope optics.
This approach not only demonstrates the successful application of modern spectroscopic techniques to environmental monitoring, expands the accessibility of spectral analysis to a broader range of researchers, but also enhances our understanding of the complex interplay between artificial lighting and its effects on the night sky.

By conducting a qualitative analysis of more than 12,000 individual zenith sky spectra collected from various locations throughout Hong Kong over multiple years, we identified the dominant influences on urban night skies. Our findings indicate that these urban environments are predominantly shaped by high-pressure sodium and compact fluorescent lamp emissions, which are widely used for road and public area illumination.
We also detected emissions from light-emitting diode, metal halide, and neon sources from the spectra. These emissions are primarily associated with facade lighting, floodlighting, billboards, and television displays, reflecting customized lighting schedules designed for business and commercial activities in urban settings.
In contrast, the skyglow in suburban areas exhibited a markedly different spectral composition. It was mainly attributed to essential lighting sources, lacking the characteristic emissions from light-emitting diode and neon sources that are typically associated with urban environments. This distinction underscores the varying impact of artificial lighting on night skies, with urban areas experiencing more complex light pollution profiles due to the diverse range of lighting technologies in use.

Previous photometric studies have indicated that late-night light-polluted skies tend to be darker, a phenomenon often attributed to the decline in human activity as the night progresses. However, the specific types of external lighting that contribute to this dimming effect have not been well understood.
In our study, we have illuminated this mystery by qualitatively and quantitatively analyzing the intensity variations of specific spectral bands associated with major ALAN sources before and after midnight. By employing a comprehensive approach to examining the spectral contributions from different lighting technologies, we were able to discern the dynamics of night sky brightness and identify which lighting sources are primarily responsible for changes in illumination levels during the late night hours.

A wealth of evidence strongly suggests that the LED, MH, and Ne spectral bands experience significant weakening in urban areas as the night progresses. This decline can be attributed to the shutdown of decorative, promotional, or advertising lighting that uses these technologies, such as billboards and display panels, which typically cease operation once the stores close for the night.
This observation is likely influenced by the successful implementation of the switching-off requirements outlined in the \textit{Charter on External Lighting}~\cite{law:2024}, which recommends specific switching-off times at 22:00, 23:00, and midnight for various types of external lighting. As a result, the diminishing brightness associated with the emissions of light-emitting diode, metal halide and neon sources during the late-night hours reflects a conscious effort to mitigate light pollution and conserve energy, thereby improving the quality of the night sky.

In contrast, our observations reveal that the intensity of the HPS and CFL spectral bands remains consistent throughout the night. This consistency indicates another common pattern of lighting use in the city, as high-pressure sodium and compact fluorescent fixtures are typically used to illuminate essential public areas such as corridors, walkways, and road networks. Unlike decorative lighting, these types of lighting are designed to remain operational throughout the night to ensure safety and accessibility for pedestrians and vehicles alike.

To our knowledge, the current work represents the first extensive spectroscopic study of the night sky in a light-polluted urban environment. This study encompasses a broad range of temporal scopes, allowing for a comprehensive analysis of nocturnal lighting dynamics. Our investigation includes decade-long analyses (Section~\ref{sec:result_decade}), yearly assessments (Section~\ref{sec:result_typical}), nightly variations (Section~\ref{sec:result_R_statistic}), and minute-by-minute evaluations (Section~\ref{sec:result_casestudies}).

The Hong Kong government initiated the retrofit of light-emitting diode street lamps between 2017 and 2018~\cite{highways_led}. In Section~\ref{sec:result_decade}, we observed that LED emissions were absent in the 2012 dataset but were clearly present in the urban center dataset from 2022 to 2023. This timing is in line with the start of the retrofit program, which highlights its substantial impact on urban lighting.
Furthermore, it is apparent that the transition to solid-state technology is still ongoing, as non-solid-state lighting fixtures continue to dominate in various areas. 
Our observations are consistent with findings from Madrid, where~\cite{robles:2021} documented the emergence of a characteristic blue bump in the sky spectrum, specifically between 442-461 nm, associated with solid-state lighting (3000K) in 2019. This change is particularly striking compared to the spectrum recorded in 2014 (see their fig. 4), coinciding with the major streetlight retrofit that began in Madrid in 2015.
In contrast, emissions from non-solid-state sources, such as mercury vapor, remained prevalent even after the retrofit of street lighting. These findings emphasize the gradual, yet impactful transition in urban settings. They also suggest that continued monitoring and analysis will be essential as this transition evolves, providing critical insights into the long-term effects of lighting changes on night sky quality and urban ecosystems.

The current study has two main limitations that should be acknowledged. 

The first limitation pertains to the data quality observed in the dark suburban locations. As illustrated in the average spectra shown in Fig.~\ref{fig:spectral_typical}, the spectroscopic signals recorded in these dark environments exhibited significant noise. This issue persisted even when using a higher-sensitivity spectrometer and averaging signals over multiple frames.
To improve the quantity and quality of spectroscopic data obtained from these dark sites -- data that could reveal not only ALAN signatures but also atmospheric emissions from natural sources -- advancements in instrumentation would be necessary.
One potential solution is to attach the spectrometer to a telescope, which would enhance light collection capabilities and improve S/N. The recent upgrades to the Hong Kong Space Museum's iObservatory telescope and spectrometer, which have improved sensitivity and resolution, represent a step in the right direction.
The deployment of specialized all-sky spectrograph devices~\cite{chernouss:2008,trondsen:2024} offers another viable option to enhance spectroscopic data collection. These advanced instruments are designed to capture fainter emissions, thereby enabling more in-depth analysis at the darkest monitoring locations.

The second limitation of this study is that it has not yet explored the influence of other less dominant lighting technologies, such as low-pressure sodium, mercury vapor, and incandescent lighting, on sky spectra. The decision to exclude these sources was primarily based on the observation that their strong spectral features were not prominently visible within the specific spectral range analyzed in this work.
However, the absence of visible features does not imply that these lighting technologies do not exert any influence on the overall sky spectra. Future studies may need to take a more comprehensive approach, specifically design to identify and quantify the traces of less prominent, yet still distinguishable, emissions from these additional lighting sources.

\section{Conclusions}\label{sec:conclusions}
In conclusion, we have demonstrated that the night sky can be studied spectroscopically, akin to an astronomical object, by examining it both temporally and spatially. This study highlights the importance of accounting for temporal variations and natural factors when interpreting the spectral composition of the sky.
The sky spectra provide a wealth of information that not only reveals natural phenomena but also highlights the impacts of human activity on nocturnal environments. Our findings emphasize the essential role of sky spectroscopy in complementing aggregate measurements typically offered by photometers and satellite sensors, providing a more nuanced understanding of light pollution and its contributing factors.
However, this work represents only a preliminary exploration of the extensive dataset we have obtained, indicating that there is much more to uncover. Future research will focus on more quantitative analyses, including assessments of changes in sky color, dependencies of band intensities, and the spectral compositions of the sky concerning various factors, such as seasonal variations, broadband brightness levels, and atmospheric conditions.
The insights gained from sky spectroscopy will enhance our understanding of the evolving lighting landscape and its implications for skyglow and the broader ecosystem. 

Our findings unequivocally illustrate the detrimental impact of ALAN on the environment. To foster sustainable urban ecosystems for both current and future generations, the implementation of targeted measures is essential. Given that we can observe ALAN spectral features from the sky, employing shielding techniques to prevent skyglow stands out as a proven strategy for significantly mitigating light pollution.
Furthermore, our observations indicate a decrease in the intensity of ALAN before midnight, suggesting that limiting the duration of ALAN operation could be an effective approach to reducing its impact. Specifically, simply turning off non-essential external lighting, including signboards, billboards, and floodlights, earlier in the evening would likely produce a positive effect.
By adopting these measures, we not only protect the natural darkness of the night skies for astronomical discoveries but also support the health of ecosystems, which benefits all living beings. Implementing such changes can improve the quality of life in urban areas, promote biodiversity, and contribute to a greater awareness of the importance of preserving our nocturnal environments. Ultimately, a collective commitment to managing light pollution will enable us to achieve a more harmonious balance between urban development and dark sky conservation.

\section*{CRediT authorship contribution statement}
\textbf{C.W. So}: Conceptualization, Data curation, Formal analysis, Funding acquisition, Investigation, Methodology, Project administration, Software, Validation, Visualization, Writing – original draft. 
\textbf{J.C.S. Pun}: Conceptualization, Funding acquisition, Methodology, Project administration, Supervision, Writing – review \& editing. 
\textbf{S. Liu}: Conceptualization, Investigation, Visualization, Writing – review \& editing.

\section*{Declaration of competing interest}
The authors declare that they have no known competing financial interests or personal relationships that could have appeared to influence the work reported in this paper.

\section*{Acknowledgments}
This research received funding from the Environment and Conservation Fund (Project IDs: 125/2018, 113/2022) of The Government of the Hong Kong Special Administrative Region. Any opinions, findings, conclusions or recommendations expressed in this article do not necessarily reflect the views of the Government of the Hong Kong Special Administrative Region and the Environment and Conservation Fund.
The \textit{Globe at Night - Sky Brightness Monitoring Network} (GaN-MN) is coorganized by the Office for Astronomy Outreach of the International Astronomy Union (IAU-OAO), The University of Hong Kong (HKU), National Astronomical Observatory of Japan (NAOJ) and Globe at Night. 
The Globe at Night citizen science campaign to monitor light pollution levels worldwide is hosted by the U.S. National Science Foundation National Optical-Infrared Astronomy Research Laboratory (NSF NOIRLab).
GaN-MN was funded by the HKU Knowledge Exchange Fund granted by the University Grants Committee (Project No.: KE-IP-2014/15-57, KE-IP-2015/16-54, KE-IP-2016/17-44, KE-IP-2017/18-54, KE-IP-2018/19-68, KE-IP-2019/20-54 and KE-IP-2020/21-78), IAU-OAO and NAOJ. 
The authors acknowledge the Hong Kong Observatory for providing the weather photos.

\section*{Data availability}
All raw data required for replication are available at \url{https://doi.org/10.25442/hku.29145164}

\bibliography{AAstyle.bib} 
\bibliographystyle{plainnat}

\end{document}